# RF engineering basic concepts: S-parameters

*F. Caspers*
CERN, Geneva, Switzerland

**Abstract**
The concept of describing RF circuits in terms of waves is discussed and the S-matrix and related matrices are defined. The signal flow graph (SFG) is introduced as a graphical means to visualize how waves propagate in an RF network. The properties of the most relevant passive RF devices (hybrids, couplers, non-reciprocal elements, etc.) are delineated and the corresponding S-parameters are given. For microwave integrated circuits (MICs) planar transmission lines such as the microstrip line have become very important.

## 1 S-parameters

The abbreviation S has been derived from the word *scattering*. For high frequencies, it is convenient to describe a given network in terms of *waves* rather than voltages or currents. This permits an easier definition of reference planes. For practical reasons, the description in terms of in- and outgoing waves has been introduced. Now, a 4-pole network becomes a 2-port and a 2n-pole becomes an n-port. In the case of an odd pole number (e.g., 3-pole), a common reference point may be chosen, attributing one pole equally to two ports. Then a 3-pole is converted into a (3+1) pole corresponding to a 2-port. As a general conversion rule, for an odd pole number one more pole is added.

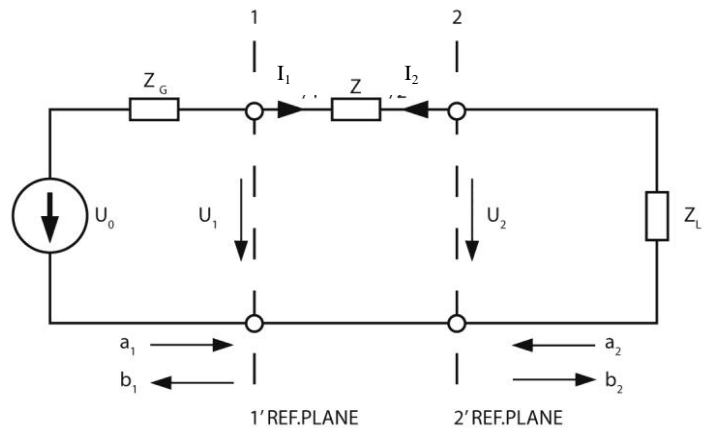

**Fig. 1:** Example for a 2-port network: a series impedance $Z$

Let us start by considering a simple 2-port network consisting of a single impedance $Z$ connected in series (Fig. 1). The generator and load impedances are $Z_G$ and $Z_L$, respectively. If $Z = 0$ and $Z_L = Z_G$ (for real $Z_G$) we have a matched load, i.e., maximum available power goes into the load and $U_1 = U_2 = U_0/2$. Note that all the voltages and currents are peak values. The lines connecting the different elements are supposed to have zero electrical length. Connections with a finite electrical length are drawn as double lines or as heavy lines. Now we would like to relate $U_0$, $U_1$ and $U_2$ to $a$ and $b$.

## 1.1 Definition of 'power waves'

The *waves* going *towards* the n-port are $\mathbf{a} = (a_1, a_2, ..., a_n)$, the *waves* travelling *away* from the n-port are $\mathbf{b} = (b_1, b_2, ..., b_n)$. By definition currents going *into* the n-port are counted positively and currents flowing out of the n-port negatively. The wave $a_1$ going into the n-port at port 1 is derived from the voltage wave going into a matched load.

In order to make the definitions consistent with the conservation of energy, the voltage is normalized to $\sqrt{Z_0}$. $Z_0$ is in general an arbitrary reference impedance, but usually the characteristic impedance of a line (e.g., $Z_0 = 50\ \Omega$) is used and very often $Z_G = Z_L = Z_0$. In the following we assume $Z_0$ to be real. The definitions of the waves $a_1$ and $b_1$ are

$$a_1 = \frac{U_0}{2\sqrt{Z_0}} = \frac{\text{incident voltage wave (port 1)}}{\sqrt{Z_0}} = \frac{U_1^{\text{inc}}}{\sqrt{Z_0}}$$

$$b_1 = \frac{U_1^{\text{refl}}}{\sqrt{Z_0}} = \frac{\text{reflected voltage wave (port 1)}}{\sqrt{Z_0}}$$

(1.1)

Note that $\mathbf{a}$ and $\mathbf{b}$ have the dimension $\sqrt{\text{power}}$ [1].

The power travelling towards port 1, $P_1^{\text{inc}}$, is simply the available power from the source, while the power coming out of port 1, $P_1^{\text{refl}}$, is given by the reflected voltage wave.

$$P_1^{\text{inc}} = \frac{1}{2}|a_1|^2 = \frac{|U_1^{\text{inc}}|^2}{2Z_0} = \frac{|I_1^{\text{inc}}|^2}{2}Z_0$$

$$P_1^{\text{refl}} = \frac{1}{2}|b_1|^2 = \frac{|U_1^{\text{refl}}|^2}{2Z_0} = \frac{|I_1^{\text{refl}}|^2}{2}Z_0$$

(1.2)

Note the factor 2 in the denominator, which comes from the definition of the voltages and currents as peak values ('European definition'). In the 'US definition' effective values are used and the factor 2 is not present, so for power calculations it is important to check how the voltages are defined. For most applications, this difference does not play a role since ratios of waves are used.

In the case of a mismatched load $Z_L$ there will be some power reflected towards (!) the 2-port from $Z_L$, leading to some incident power at port 2.

$$P_2^{\text{inc}} = \frac{1}{2}|a_2|^2 \ .$$

(1.3)

There is also the outgoing wave of port 2 which may be considered as the superimposition of a wave that has gone through the 2-port from the generator and a reflected part from the mismatched load. We have defined $a_1 = U_0/(2\sqrt{Z_0}) = U^{\text{inc}}/\sqrt{Z_0}$ with the incident voltage wave $U^{\text{inc}}$. In analogy to that we can also quote for the power wave $a_1 = I^{\text{inc}}\sqrt{Z_0}$ with the incident current wave $I^{\text{inc}}$. We obtain the *general definition* of the waves $a_i$ travelling into and $b_i$ travelling *out of* an n-port:

$$a_i = \frac{U_i + I_i Z_0}{2\sqrt{Z_0}}$$

$$b_i = \frac{U_i - I_i Z_0}{2\sqrt{Z_0}} \tag{1.4}$$

Solving these two equations, $U_i$ and $I_i$ can be obtained for a given $a_i$ and $b_i$ as

$$U_i = \sqrt{Z_0}\,(a_i + b_i) = U_i^{\text{inc}} + U_i^{\text{refl}}$$

$$I_i = \frac{1}{\sqrt{Z_0}}(a_i - b_i) = \frac{U_i^{\text{refl}}}{Z_0}. \tag{1.5}$$

For a harmonic excitation $u(t) = \text{Re}\{U e^{j\omega t}\}$ the power going *into* port $i$ is given by

$$P_i = \frac{1}{2}\text{Re}\{U_i I_i^*\}$$

$$P_i = \frac{1}{2}\text{Re}\left\{\left(a_i a_i^* - b_i b_i^*\right) + \left(a_i^* b_i - a_i b_i^*\right)\right\} \tag{1.6}$$

$$P_i = \frac{1}{2}\left(a_i a_i^* - b_i b_i^*\right)$$

The term $(a_i^* b_i - a_i b_i^*)$ is a purely imaginary number and vanishes when the real part is taken.

## 1.2  The S-matrix

The relation between $a_i$ and $b_i$ ($i = 1...n$) can be written as a system of $n$ linear equations ($a_i$ being the independent variable, $b_i$ the dependent variable)

$$b_1 = S_{11} a_1 + S_{12} a_2$$

$$b_2 = S_{21} a_1 + S_{22} a_2 \tag{1.7}$$

or, in matrix formulation

$$\mathbf{b} = \mathbf{S}\,\mathbf{a} \tag{1.8}$$

The physical meaning of $S_{11}$ is the input reflection coefficient with the output of the network terminated by a matched load ($a_2 = 0$). $S_{21}$ is the forward transmission (from port 1 to port 2), $S_{12}$ the reverse transmission (from port 2 to port 1) and $S_{22}$ the output reflection coefficient.

When measuring the S-parameter of an $n$-port, *all n* ports must be terminated by a matched load (not necessarily equal value for all ports), including the port connected to the generator (matched generator).

Using Eqs. (1.4) and (1.7) we find the reflection coefficient of a single impedance $Z_L$ connected to a generator of source impedance $Z_0$ (Fig. 1, case $Z_G = Z_0$ and $Z = 0$):

$$S_{11} = \left.\frac{b_1}{a_1}\right|_{a_2=0} = \frac{U_1 - I_1 Z_0}{U_1 + I_1 Z_0} = \frac{Z_L - Z_0}{Z_L + Z_0} = \Gamma = \frac{(Z_L/Z_0) - 1}{(Z_L/Z_0) + 1} \tag{1.9}$$

which is the familiar formula for the reflection coefficient $\Gamma$ (sometimes also denoted as $\rho$).

Let us now determine the S-parameters of the impedance $Z$ in Fig. 1, assuming again $Z_G = Z_L = Z_0$. From the definition of $S_{11}$ we have

$$S_{11} = \frac{b_1}{a_1} = \frac{U_1 - I_1 Z_0}{U_1 + I_1 Z_0}$$

$$U_1 = U_0 \frac{Z_0 + Z}{2Z_0 + Z}, \quad U_2 = U_0 \frac{Z_0}{2Z_0 + Z}, \quad I_1 = \frac{U_0}{2Z_0 + Z} = -I_2$$

$$\Rightarrow S_{11} = \frac{Z}{2Z_0 + Z} \tag{1.10}$$

and in a similar fashion we get

$$S_{21} = \frac{b_2}{a_1} = \frac{U_2 - I_2 Z_0}{U_1 + I_1 Z_0} = \frac{2Z_0}{2Z_0 + Z}. \tag{1.11}$$

Due to the symmetry of the element $S_{22} = S_{22}$ and $S_{12} = S_{21}$. Please note that for this case we obtain $S_{11} + S_{21} = 1$. The full S-matrix of the element is then

$$\mathbf{S} = \begin{pmatrix} \dfrac{Z}{2Z_0 + Z} & \dfrac{Z_0 + Z}{2Z_0 + Z} \\ \dfrac{Z_0 + Z}{2Z_0 + Z} & \dfrac{Z}{2Z_0 + Z} \end{pmatrix}. \tag{1.12}$$

### 1.3 The transfer matrix

The S-matrix introduced in the previous section is a very convenient way to describe an *n*-port in terms of waves. It is very well adapted to measurements. However, it is not well suited for characterizing the response of a number of cascaded 2-ports. A very straightforward manner for the problem is possible with the T-matrix (transfer matrix), which directly relates the waves on the input and on the output [2]

$$\begin{pmatrix} b_1 \\ a_1 \end{pmatrix} = \begin{pmatrix} T_{11} & T_{12} \\ T_{21} & T_{22} \end{pmatrix} \begin{pmatrix} a_2 \\ b_2 \end{pmatrix}. \tag{1.13}$$

The conversion formulae between the S and T matrix are given in Appendix I. While the S-matrix exists for any 2-port, in certain cases, e.g., no transmission between port 1 and port 2, the T-matrix is not defined. The T-matrix $\mathbf{T_M}$ of *m* cascaded 2-ports is given by (as in Refs. [2, 3]):

$$\mathbf{T_M} = \mathbf{T_1 T_2 \ldots T_m}. \tag{1.14}$$

Note that in the literature different definitions of the T-matrix can be found and the individual matrix elements depend on the definition used.

### 2 Properties of the S-matrix of an *n*-port

A generalized *n*-port has $n^2$ scattering coefficients. While the $S_{ij}$ may be all independent, in general due to symmetries etc. the number of independent coefficients is much smaller.

- An *n*-port is *reciprocal* when $S_{ij} = S_{ji}$ for all $i$ and $j$. Most passive components are reciprocal (resistors, capacitors, transformers, etc., except for structures involving magnetized ferrites, plasmas etc.), active components such as amplifiers are generally non-reciprocal.

- A two-port is *symmetric*, when it is reciprocal ($S_{21} = S_{12}$) and when the input and output reflection coefficients are equal ($S_{22} = S_{11}$).
- An N-port is *passive and lossless* if its **S** matrix is *unitary*, i.e., $\mathbf{S}^\dagger \mathbf{S} = \mathbf{1}$, where $\mathbf{x}^\dagger = (\mathbf{x}^*)^T$ is the conjugate transpose of **x**. For a two-port this means

$$(S^*)^T S = \begin{pmatrix} S_{11}^* & S_{21}^* \\ S_{12}^* & S_{22}^* \end{pmatrix} \begin{pmatrix} S_{11} & S_{12} \\ S_{21} & S_{22} \end{pmatrix} = \begin{pmatrix} 1 & 0 \\ 0 & 1 \end{pmatrix}$$

(2.1)

which yields three conditions

$$|S_{11}|^2 + |S_{21}|^2 = 1$$
$$|S_{12}|^2 + |S_{22}|^2 = 1$$

(2.2)

$$S_{11}^* S_{12} + S_{21}^* S_{22} = 0$$

(2.3)

Splitting up the last equation in the modulus and argument yields

$$|S_{11}||S_{12}| = |S_{21}||S_{22}| \quad \text{and}$$

(2.4)

$$-\arg S_{11} + \arg S_{12} = -\arg S_{21} + \arg S_{22} + \pi$$

where arg(*x*) is the argument (angle) of the complex variable *x*. Combining Eq. (2.2) with the first of Eq. (2.4) then gives

$$|S_{11}| = |S_{22}|, \quad |S_{12}| = |S_{21}|$$
$$|S_{11}| = \sqrt{1 - |S_{12}|^2}.$$

(2.5)

Thus any lossless 2-port can be characterized by one modulus and three angles.

In general the S-parameters are complex and frequency dependent. Their phases change when the reference plane is moved. Often the S-parameters can be determined from considering symmetries and, in case of lossless networks, energy conservation.

## 2.1 Examples of S-matrices

*1-port*

- Ideal short $\qquad S_{11} = -1$
- Ideal termination $\qquad S_{11} = 0$
- Active termination (reflection amplifier) $\qquad |S_{11}| > 1$

*2-port*

- Ideal transmission line of length *l*

$$\mathbf{S} = \begin{pmatrix} 0 & e^{-\gamma l} \\ e^{-\gamma l} & 0 \end{pmatrix}$$

where $\gamma = \alpha + j\beta$ is the complex propagation constant, $\alpha$ the line attenuation in [neper/m] and $\beta = 2\pi/\lambda$ with the wavelength $\lambda$. For a lossless line we obtain $|S_{21}| = 1$.

- Ideal phase shifter

$$\mathbf{S} = \begin{pmatrix} 0 & e^{-j\phi_{12}} \\ e^{-j\phi_{21}} & 0 \end{pmatrix}$$

For a reciprocal phase shifter $\varphi_{12} = \varphi_{21}$, while for the gyrator $\varphi_{12} = \varphi_{21} + \pi$. An ideal gyrator is lossless ($\mathbf{S}^{\dagger}\mathbf{S} = \mathbf{1}$), but it is not reciprocal. Gyrators are often implemented using active electronic components, however, in the microwave range, passive gyrators can be realized using magnetically saturated ferrite elements.

- Ideal, reciprocal attenuator

$$\mathbf{S} = \begin{pmatrix} 0 & e^{-\alpha} \\ e^{-\alpha} & 0 \end{pmatrix}$$

with the attenuation $\alpha$ in neper. The attenuation in decibel is given by $A = -20*\log_{10}(S_{21})$, 1 Np = 8.686 dB. An attenuator can be realized, for example, with three resistors in a T circuit. The values of the required resistors are

$$R_1 = Z_0 \frac{k-1}{k+1}$$

$$R_2 = Z_0 \frac{2k}{k^2-1}$$

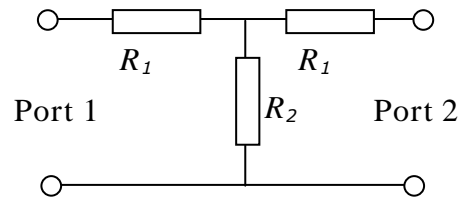

where $k$ is the voltage attenuation factor and $Z_0$ the reference impedance, e.g., 50 Ω.

- Ideal isolator

$$\mathbf{S} = \begin{pmatrix} 0 & 0 \\ 1 & 0 \end{pmatrix}$$

The isolator allows transmission in one direction only; it is used, for example, to avoid reflections from a load back to the generator.

- Ideal amplifier

$$\mathbf{S} = \begin{pmatrix} 0 & 0 \\ G & 0 \end{pmatrix}$$

with the gain $G > 1$.

*3-port*

Several types of 3-port are in use, e.g., power dividers, circulators, T junctions. It can be shown that a 3-port cannot be lossless, reciprocal, and matched at all three ports at the same time. The following three components have two of the above characteristics:

- Resistive power divider: It consists of a resistor network and is reciprocal, matched at all ports but lossy. It can be realized with three resistors in a triangle configuration. With port 3 connected to ground, the resulting circuit is similar to a 2-port attenuator but no longer matched at port 1 and port 2.

$$S = \begin{pmatrix} 0 & 1/2 & 1/2 \\ 1/2 & 0 & 1/2 \\ 1/2 & 1/2 & 0 \end{pmatrix}$$

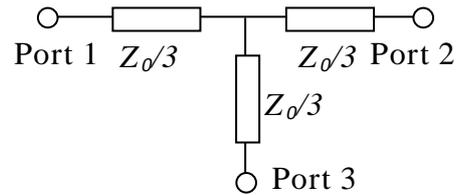

- The T splitter is reciprocal and lossless but not matched at all ports.

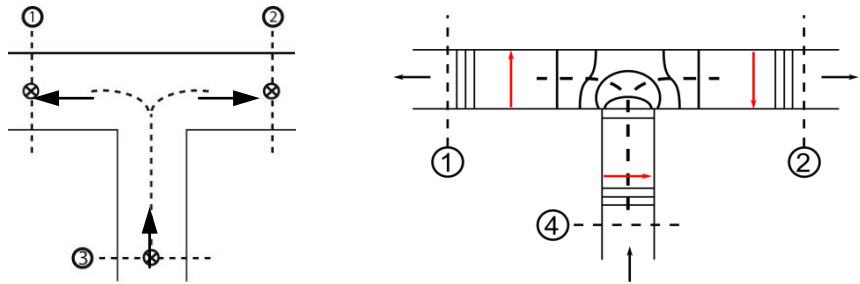

**Fig. 2:** The two versions of the $H_{10}$ waveguide T splitter: H-plane and E-plane splitter

Using the fact that the 3-port is lossless and there are symmetry considerations one finds, for appropriate reference planes for H- and E- plane splitters

$$S_H = \frac{1}{2}\begin{pmatrix} 1 & -1 & \sqrt{2} \\ -1 & 1 & \sqrt{2} \\ \sqrt{2} & \sqrt{2} & 0 \end{pmatrix} \qquad S_E = \frac{1}{2}\begin{pmatrix} 1 & 1 & \sqrt{2} \\ 1 & 1 & -\sqrt{2} \\ \sqrt{2} & -\sqrt{2} & 0 \end{pmatrix}.$$

- The ideal circulator is lossless, matched at all ports, but not reciprocal. A signal entering the ideal circulator at one port is transmitted *exclusively* to the next port in the sense of the arrow (Fig. 3).

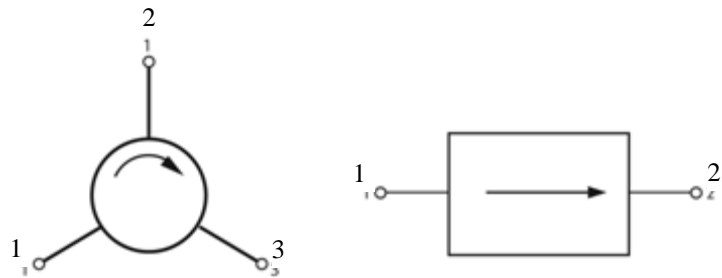

**Fig. 3:** 3-port circulator and 2-port isolator. The circulator can be converted into isolator by connecting a matched load to port 3.

Accordingly, the S-matrix of the isolator has the following form:

$$S = \begin{pmatrix} 0 & 0 & 1 \\ 1 & 0 & 0 \\ 0 & 1 & 0 \end{pmatrix}.$$

When port 3 of the circulator is terminated with a matched load we get a two-port called isolator, which lets power pass only from port 1 to port 2 (see section about 2-ports). A circulator, like the gyrator and other passive non-reciprocal elements contains a volume of ferrite. This ferrite is normally magnetized into saturation by an external magnetic field. The magnetic properties of a saturated RF ferrite have to be characterized by a µ-tensor. The real and imaginary part of each complex element µ are µ′ and µ″. They are strongly dependent on the bias field. The $\mu_+$ and $\mu_-$ represent the permeability seen by a right- and left-hand circular polarized wave traversing the ferrite (Fig. 4). In this figure the µ′ and the µ″ for the right- and left-hand circular polarized waves are depicted denoted as $\mu'_+$ and $\mu'_-$ for the real part of µ and correspondingly for the imaginary part.

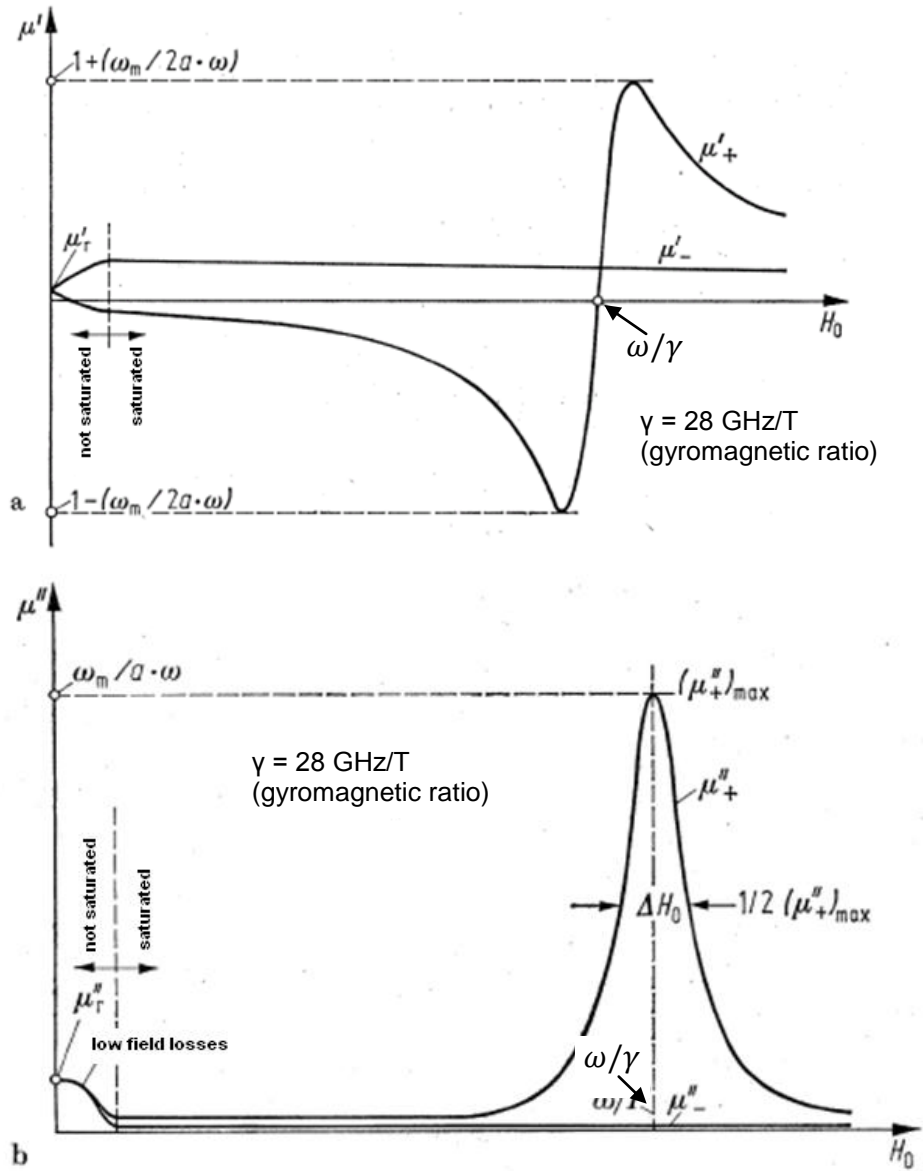

**Fig. 4:** Real part $\mu_r'$ and imaginary part $\mu_r''$ of the complex permeability µ. The right- and left-hand circularly polarized waves in a microwave ferrite are $\mu_+$ and $\mu_-$ respectively. At the gyromagnetic resonance the right-hand polarized has high losses, as can be seen from the peak in the lower image.

In Figs. 5 and 6 practical implementations of circulators are shown. The magnetically polarized ferrite provides the required nonreciprocal properties. As a result, power is only transmitted from port 1 to port 2, from port 2 to port 3 and from port 3 to port 1. A detailed discussion of the different working principles of circulators can be found in the literature [2, 4].

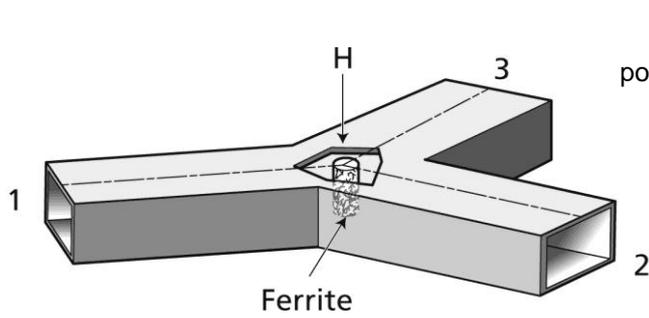 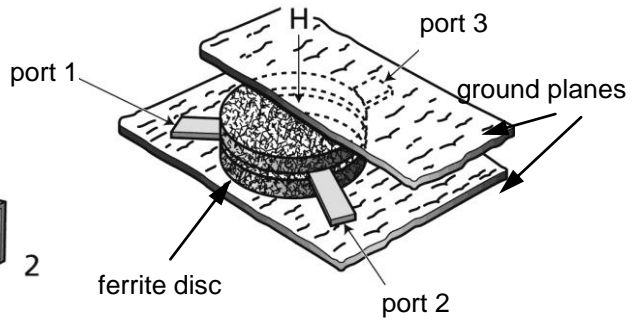

**Fig. 5:** Waveguide circulator  **Fig. 6:** Stripline circulator

The Faraday rotation isolator uses the $TE_{10}$ mode in a rectangular waveguide, which has a vertically polarized H field in the waveguide on the left (Fig. 7). After a transition to a circular waveguide, the polarization of the waveguide mode is rotated counter clockwise by 45° by a ferrite. Then follows a transition to another rectangular waveguide which is rotated by 45° such that the rotated forward wave can pass unhindered. However, a wave coming from the other side will have its polarization rotated by 45° *clockwise* as seen from the right side. In the waveguide on the left the backward wave arrives with a horizontal polarization. The horizontal attenuation foils dampen this mode, while they hardly affect the forward wave. Therefore the Faraday isolator allows transmission only from port 1 to port 2.

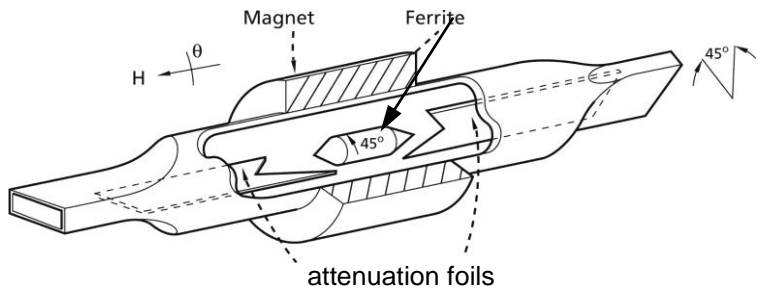

**Fig. 7:** Faraday rotation isolator

The frequency range of ferrite-based, non-reciprocal elements extends from about 50 MHz up to optical wavelengths (Faraday rotator) [4]. Finally, it should be noted that all non-reciprocal elements can be made from a combination of an ideal gyrator (non-reciprocal phase shifter) and other passive, reciprocal elements, e.g., 4-port T-hybrids or magic tees.

*The S-matrix of a 4-port*

As a first example let us consider a combination of E-plane and H-plane waveguide 'T's (Fig. 8).

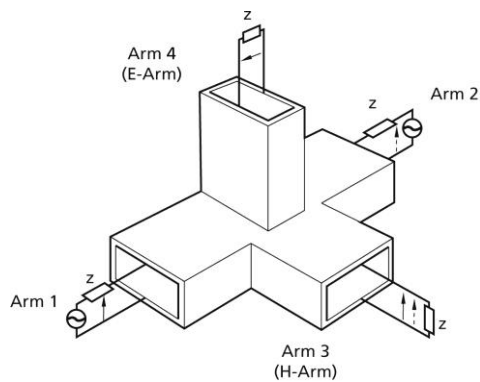

**Fig. 8:** Hybrid 'T', Magic 'T', 180° hybrid. Ideally there is no crosstalk between port 3 and port 4 nor between port 1 and port 2

This configuration is called a Magic 'T' and has the S-matrix:

$$\mathbf{S} = \frac{1}{\sqrt{2}} \begin{pmatrix} 0 & 0 & 1 & 1 \\ 0 & 0 & 1 & -1 \\ 1 & 1 & 0 & 0 \\ 1 & -1 & 0 & 0 \end{pmatrix}$$

As usual the coefficients of the S-matrix can be found by using the unitary condition and mechanical symmetries. Contrary to a 3-port, a 4-port may be lossless, reciprocal, and matched at all ports simultaneously. With a suitable choice of the reference planes the very simple S-matrix given above results.

In practice, certain measures are required to make the 'T' a 'magic' one, such as small matching stubs in the centre of the 'T'. Today, T-hybrids are often produced not in waveguide technology, but as coaxial lines and printed circuits. They are widely used for signal combination or splitting in pickups and kickers for particle accelerators. In a simple vertical-loop pickup the signal outputs of the upper and lower electrodes are connected to arm 1 and arm 2, and the sum (Σ) and difference (Δ) signals are available from the H arm and E arm, respectively. This is shown in Fig. 8 assuming two generators connected to the collinear arms of the magic T. The signal from generator 1 is split and fed with equal amplitudes into the E and H arm, which correspond to the Δ and Σ ports. The signal from generator 2 propagates in the same way. Provided both generators have equal amplitude and phase, the signals cancel at the Δ port and the sum signal shows up at the Σ port. The bandwidth of a waveguide magic 'T' is around one octave or the equivalent $H_{10}$-mode waveguide band. Broadband versions of 180° hybrids may have a frequency range from a few MHz to some GHz.

Another important element is the directional coupler. A selection of possible waveguide couplers is depicted in Fig. 9.

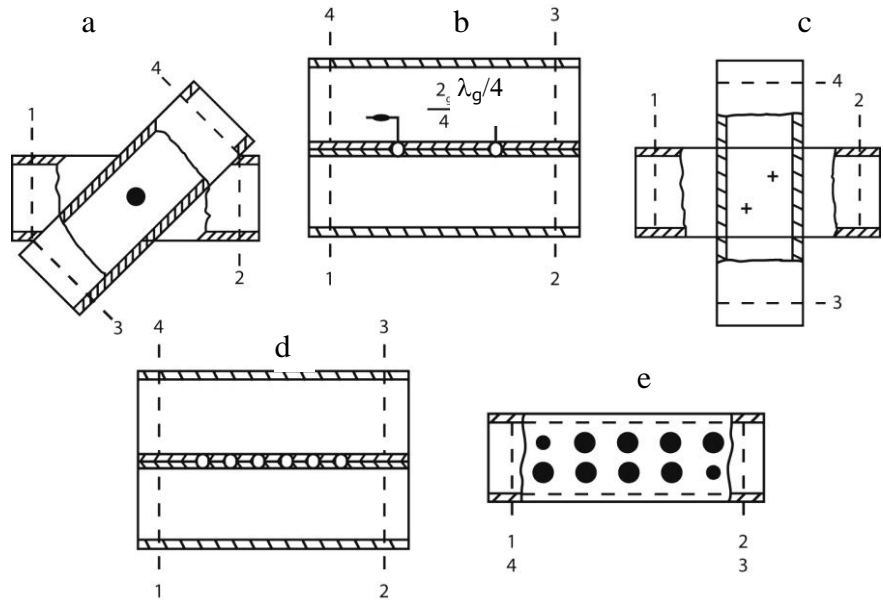

**Fig. 9:** Waveguide directional couplers: (a) single-hole, (b,c) double-hole, and (d,e) multiple-hole types

There is a common principle of operation for all directional couplers: we have two transmission lines (waveguide, coaxial line, strip line, microstrip), and coupling is adjusted such that part of the power linked to a travelling wave in line 1 excites travelling waves in line 2. The coupler is directional when the coupled energy mainly propagates in a single travelling wave, i.e., when there is no equal propagation in the two directions.

The single-hole coupler (Fig. 9), also known as a Bethe coupler, takes advantage of the electric and magnetic polarizability of a small (d<<λ) coupling hole. A propagating wave in the main line excites electric and magnetic currents in the coupling hole. Each of these currents gives rise to travelling waves in both directions. The electric coupling is independent of the angle α between the waveguides (also possible with two coaxial lines at an angle α). In order to get directionality, at least two coupling mechanisms are necessary, i.e., two coupling holes or electric and magnetic coupling. For the Bethe coupler the electric coupling does not depend on the angle α between the waveguides, while the magnetic coupling is angle-dependent. It can be shown that for α = 30° the electric and magnetic components cancel in one direction and add in the other and we have a directional coupler. The physical mechanism for the other couplers shown in Fig. 9 is similar. Each coupling hole excites waves in both directions but the superposition of the waves coming from all coupling holes leads to a preference for a particular direction.

*Example: the 2-hole, λ/4 coupler*

For a wave incident at port 1 two waves are excited at the positions of the coupling holes in line 2 (top of Fig. 9b). For a backwards coupling towards port 4 these two waves have a phase shift of 180°, so they cancel. For the forward coupling the two waves add up in phase and all the power coupled to line 2 leaves at port 3. Optimum directivity is only obtained in a narrow frequency range where the distance of the coupling holes is roughly λ/4. For larger bandwidths, multiple hole couplers are used. The holes need not be circular; they may be longitudinally or transversely orientated slots, crosses, etc.

Besides waveguide couplers there exists a family of printed circuit couplers (stripline, microstrip) and also lumped element couplers (like transformers). To characterize directional couplers, two important figures are always required, the *coupling* and the *directivity*. For the elements shown in Fig. 9, the coupling appears in the S-matrix as the coefficient

$$|S_{13}| = |S_{31}| = |S_{42}| = |S_{24}|$$

with $\alpha_c$ = -20 log$|S_{13}|$ in dB being the coupling attenuation.

The *directivity* is the ratio of the desired coupled wave to the *undesired* (i.e., wrong direction) coupled wave, e.g.,

$$\alpha_d = 20\, log\, \frac{|S_{31}|}{|S_{41}|} \quad \text{directivity} \,[\text{dB}].$$

Practical numbers for the coupling are 3 dB, 6 dB, 10 dB, and 20 dB with directivities usually better than 20 dB. Note that the ideal 3 dB coupler (like most directional couplers) often has a π/2 phase shift between the main line and the coupled line (90° hybrid). The following relations hold for an ideal directional coupler with properly chosen reference planes

$$\begin{aligned}
S_{11} &= S_{22} = S_{33} = S_{44} = 0 \\
S_{21} &= S_{12} = S_{43} = S_{34} \\
S_{31} &= S_{13} = S_{42} = S_{24} \\
S_{41} &= S_{14} = S_{32} = S_{23}
\end{aligned} \qquad (2.6)$$

$$S = \begin{pmatrix} 0 & \sqrt{1-|S_{13}|^2} & \pm j|S_{13}| & 0 \\ \sqrt{1-|S_{13}|^2} & 0 & 0 & \pm j|S_{13}| \\ \pm j|S_{13}| & 0 & 0 & \sqrt{1-|S_{13}|^2} \\ 0 & \pm j|S_{13}| & \sqrt{1-|S_{13}|^2} & 0 \end{pmatrix} \qquad (2.7)$$

and for the 3 dB coupler ($\pi/2$-hybrid)

$$S_{3dB} = \frac{1}{\sqrt{2}} \begin{pmatrix} 0 & 1 & \pm j & 0 \\ 1 & 0 & 0 & \pm j \\ \pm j & 0 & 0 & 1 \\ 0 & \pm j & 1 & 0 \end{pmatrix}. \qquad (2.8)$$

As further examples of 4-ports, the 4-port circulator and the one-to-three power divider should be mentioned.

For more general cases, one must keep in mind that a port is assigned to each waveguide or TEM-mode considered. Since for waveguides the number of propagating modes increases with frequency, a network acting as a 2-port at low frequencies will become a *2n-port* at higher frequencies (Fig. 10), with *n* increasing each time a new waveguide mode starts to propagate. Also a TEM line beyond cutoff is a multiport. In certain cases modes below cutoff may be taken into account, e.g., for calculation of the scattering properties of waveguide discontinuities, using the S-matrix approach.

There are different technologies for realizing microwave elements such as directional couplers and T-hybrids. Examples are the stripline coupler shown in Fig. 11, the 90°, 3 dB coupler in Fig. 12 and the printed-circuit magic T in Fig. 13.

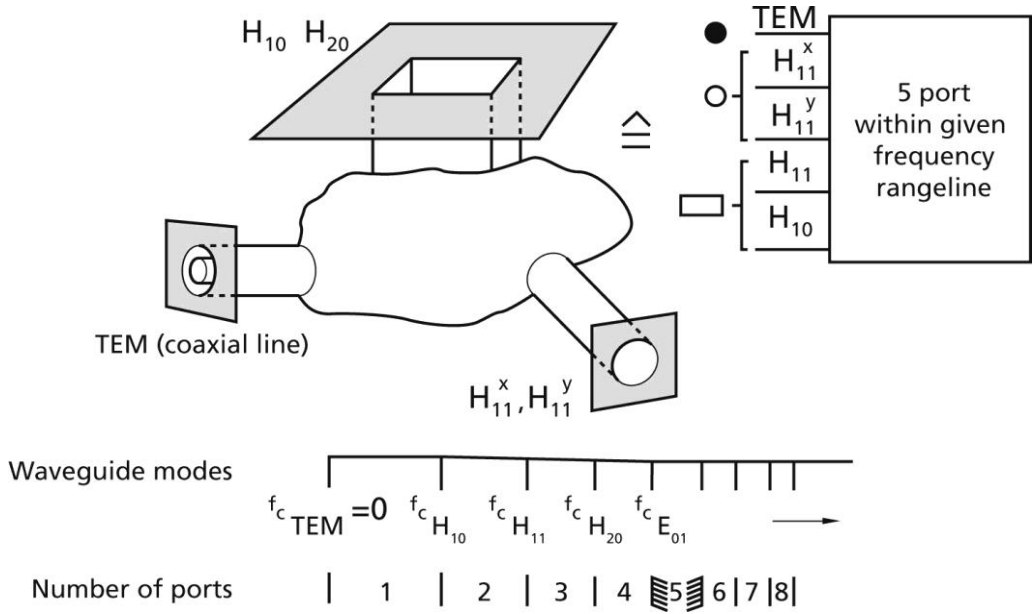

**Fig. 10:** Example of a multiport comprising waveguide ports. At higher frequencies more waveguide modes can propagate; the port number increases correspondingly.

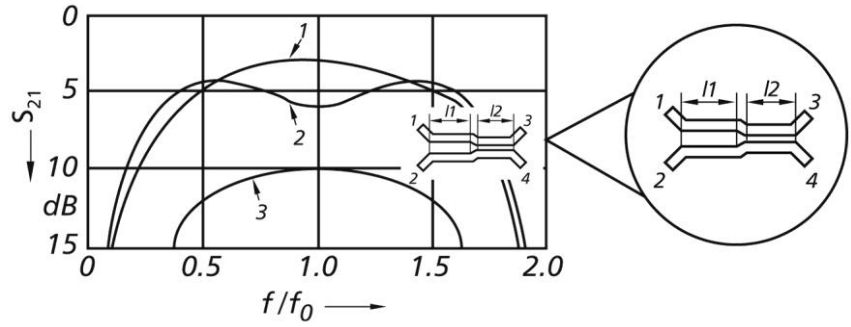

**Fig. 11**: Two-stage stripline directional coupler. curve 1: 3 dB coupler, curve 2: broadband 5 dB coupler, curve 3: 10 dB coupler (cascaded 3-dB and 10-dB coupler) [2].

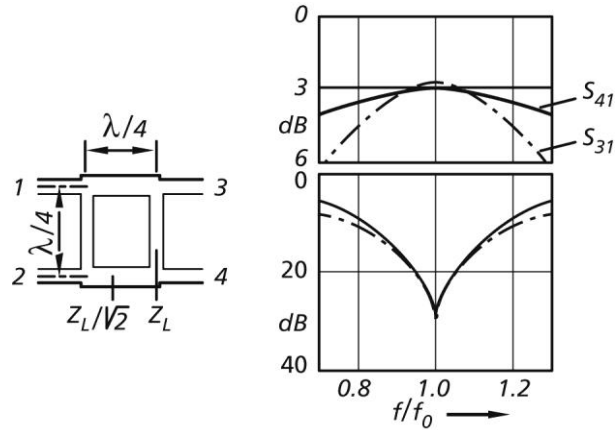

**Fig. 12:** 90° 3-dB coupler [2]

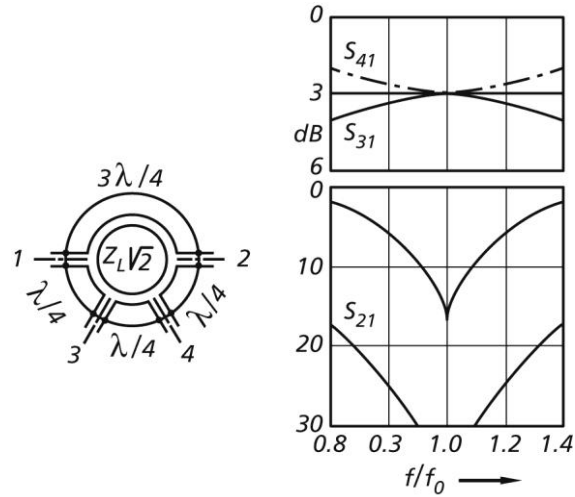

**Fig. 13:** Magic T in a printed circuit version [2]

# 3 Basic properties of striplines, microstrip and slotlines

## 3.1 Striplines

A stripline is a flat conductor between a top and bottom ground plane. The space around this conductor is filled with a homogeneous dielectric material. This line propagates a pure TEM mode. With the static capacity per unit length, $C'$, the static inductance per unit length, $L'$, the relative permittivity of the dielectric, $\varepsilon_r$, and the speed of light, $c$, the characteristic impedance $Z_0$ of the line (Fig. 14) is given by

$$Z_0 = \sqrt{\frac{L'}{C'}}$$
$$v_{ph} = \frac{c}{\sqrt{\varepsilon_r}} = \frac{1}{\sqrt{L'C'}} \,. \tag{3.1}$$
$$Z_0 = \sqrt{\varepsilon_r} \cdot \frac{1}{C'c}$$

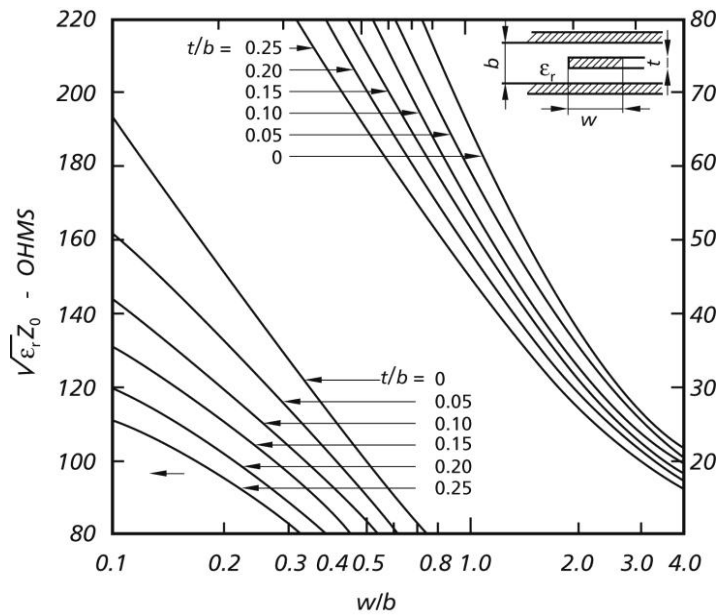

**Fig. 14:** Characteristic impedance of striplines [5]

For a mathematical treatment, the effect of the fringing fields may be described in terms of static capacities (see Fig. 15) [5]. The total capacity is the sum of the principal and fringe capacities $C_p$ and $C_f$.

$$C_{tot} = C_{p1} + C_{p2} + 2C_{f1} + 2C_{f2} \,. \tag{3.2}$$

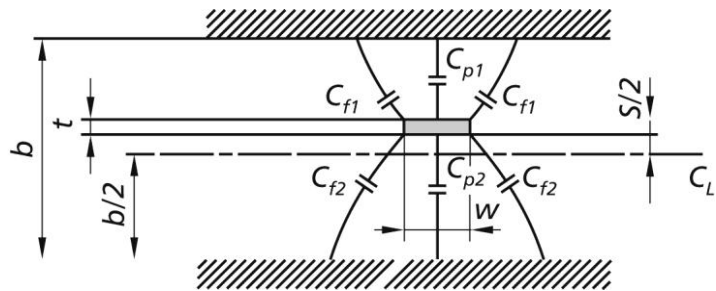

**Fig. 15:** Design, dimensions and characteristics for offset centre-conductor strip transmission line [5]

For striplines with an homogeneous dielectric the phase velocity is the same, and frequency independent, for all TEM-modes. A configuration of two coupled striplines (3-conductor system) may have two independent TEM-modes, an odd mode and an even mode (Fig. 16).

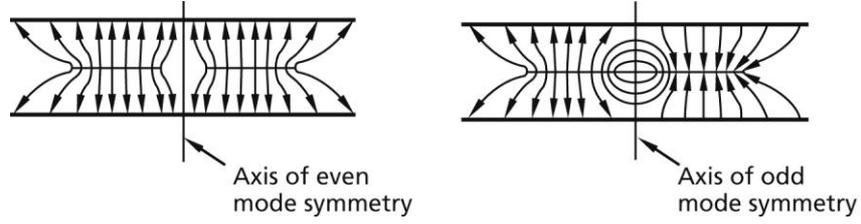

Fig. 16: Even and odd mode in coupled striplines [5]

The analysis of coupled striplines is required for the design of directional couplers. Besides the phase velocity the odd and even mode impedances $Z_{0,odd}$ and $Z_{0,even}$ must be known. They are given as a good approximation for the side coupled structure (Fig. 17, left) [5]. They are valid as a good approximation for the structure shown in Fig. 17.

$$Z_{0,even} = \frac{1}{\sqrt{\varepsilon_r}} \cdot \frac{94.15\,\Omega}{\dfrac{w}{b} + \dfrac{\ln 2}{\pi} + \dfrac{1}{\pi}\ln\left(1+\tanh\left(\dfrac{\pi}{2}\dfrac{s}{b}\right)\right)}$$

$$Z_{0,odd} = \frac{1}{\sqrt{\varepsilon_r}} \cdot \frac{94.15\,\Omega}{\dfrac{w}{b} + \dfrac{\ln 2}{\pi} + \dfrac{1}{\pi}\ln\left(1+\coth\left(\dfrac{\pi}{2}\dfrac{s}{b}\right)\right)}$$

(3.3)

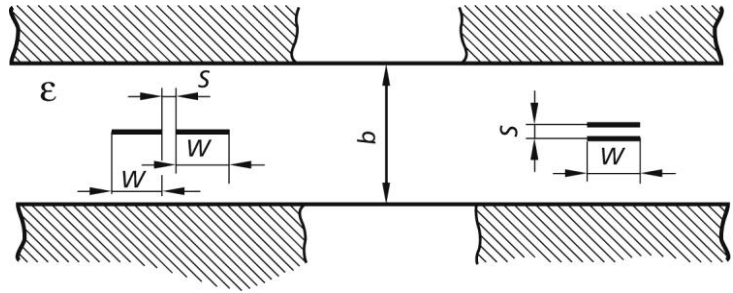

Fig. 17: Types of coupled striplines [5]: left: side coupled parallel lines, right: broad-coupled parallel lines

A graphical presentation of Eqs. (3.3) is also known as the Cohn nomographs [5]. For a quarter-wave directional coupler (single section in Fig. 11) very simple design formulae can be given.

$$Z_{0,odd} = Z_0 \sqrt{\frac{1+C_0}{1-C_0}}$$

$$Z_{0,even} = Z_0 \sqrt{\frac{1-C_0}{1+C_0}}$$

(3.4)

$$Z_0 = \sqrt{Z_{0,odd} Z_{0,even}}$$

where $C_0$ is the voltage coupling ratio of the λ/4 coupler.

In contrast to the 2-hole waveguide coupler this type couples backwards, i.e., the coupled wave leaves the coupler in the direction opposite to the incoming wave. Stripline coupler technology is now rather widespread, and very cheap high-quality elements are available in a wide frequency range. An even simpler way to make such devices is to use a section of shielded 2-wire cable.

### 3.2 Microstrip

A microstripline may be visualized as a stripline with the top cover and the top dielectric layer taken away (Fig. 18). It is thus an asymmetric open structure, and only part of its cross section is filled with a dielectric material. Since there is a transversely inhomogeneous dielectric, only a quasi-TEM wave exists. This has several implications such as a frequency-dependent characteristic impedance and a considerable dispersion (Fig. 19).

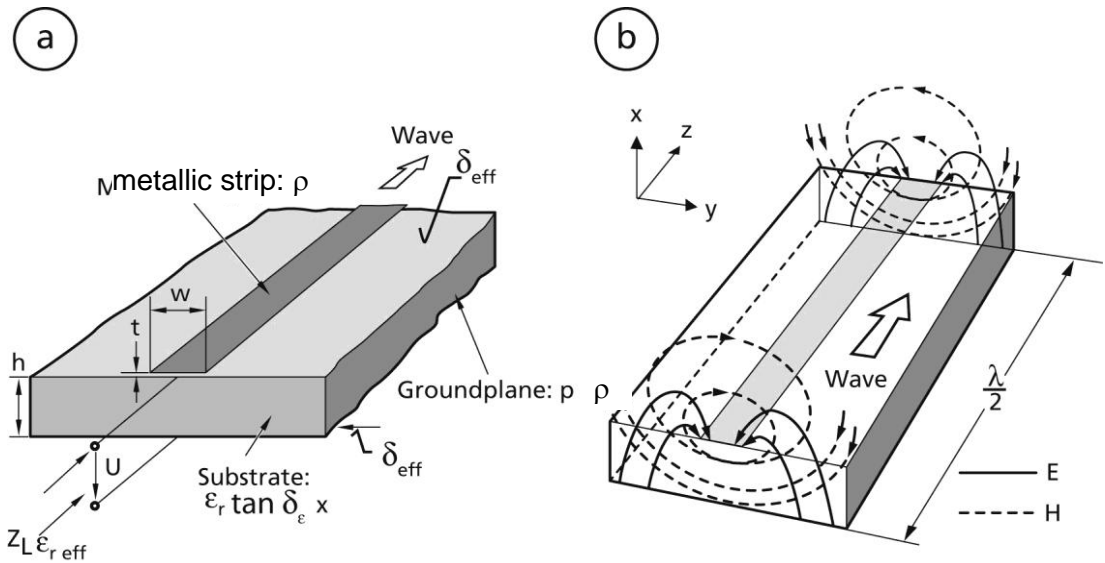

**Fig. 18:** Microstripline: a) Mechanical construction, b) Static field approximation [6]

An exact field analysis for this line is rather complicated and there exist a considerable number of books and other publications on the subject [6, 7]. Owing to dispersion of the microstrip, the calculation of coupled lines and thus the design of couplers and related structures is also more complicated than in the case of the stripline. Microstrips tend to radiate at all kind of discontinuities such as bends, changes in width, through holes, etc.

With all the above-mentioned disadvantages in mind, one may question why they are used at all. The mains reasons are the cheap production, once a conductor pattern has been defined, and easy access to the surface for the integration of active elements. Microstrip circuits are also known as Microwave Integrated Circuits (MICs). A further technological step is the MMIC (Monolithic Microwave Integrated Circuit) where active and passive elements are integrated on the same semiconductor substrate.

In Figs. 20 and 21 various planar printed transmission lines are depicted. The microstrip with overlay is relevant for MMICs and the strip dielectric wave guide is a 'printed optical fibre' for millimetre-waves and integrated optics [7].

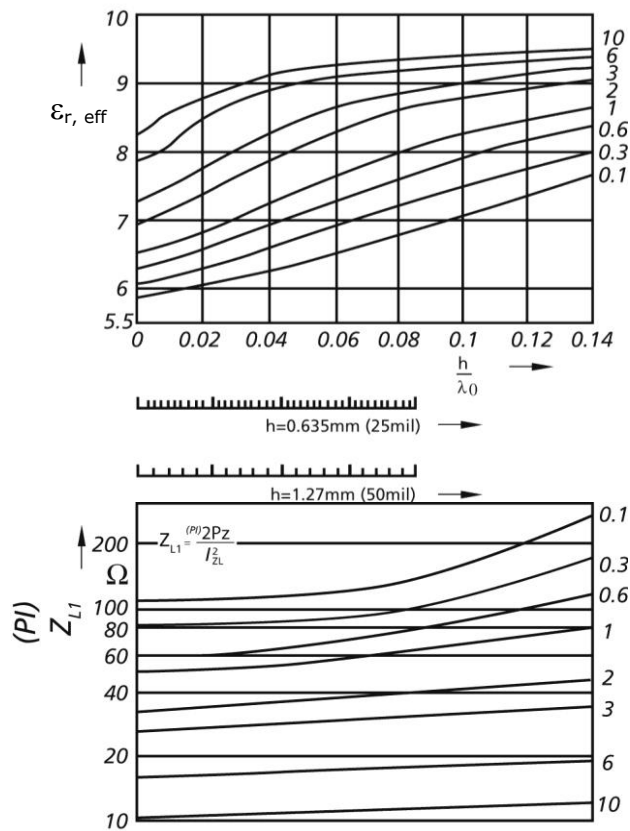

**Fig. 19:** Characteristic impedance (current/power definition) and effective permittivity of a microstrip line [6]

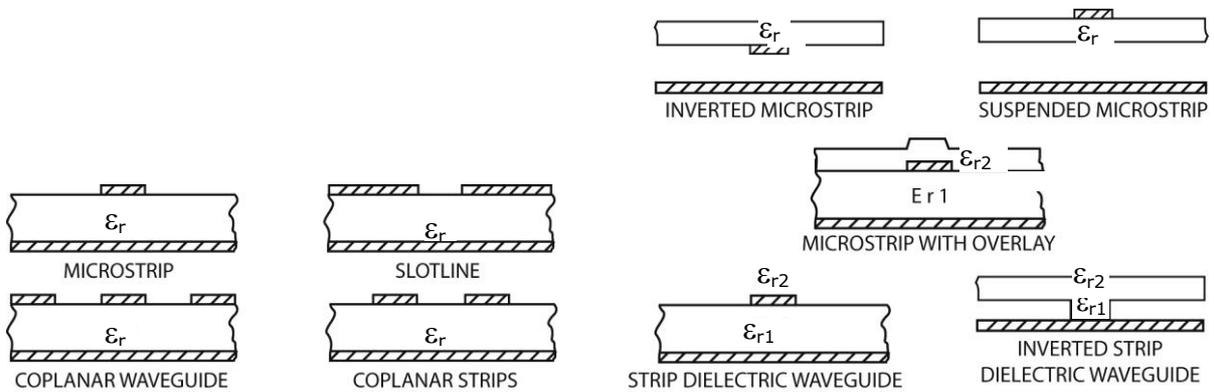

**Fig. 20:** Planar transmission lines used in MIC

**Fig. 21:** Various transmission lines derived from the microstrip concept [7]

## 3.3 Slotlines

The slotline may be considered as the dual structure of the microstrip. It is essentially a slot in the metallization of a dielectric substrate as shown in Fig. 22. The characteristic impedance and the effective dielectric constant exhibit similar dispersion properties to those of the microstrip line. A unique feature of the slotline is that it may be combined with microstrip lines on the same substrate.

This, in conjunction with through holes, permits interesting topologies such as pulse inverters in sampling heads (e.g., for sampling scopes).

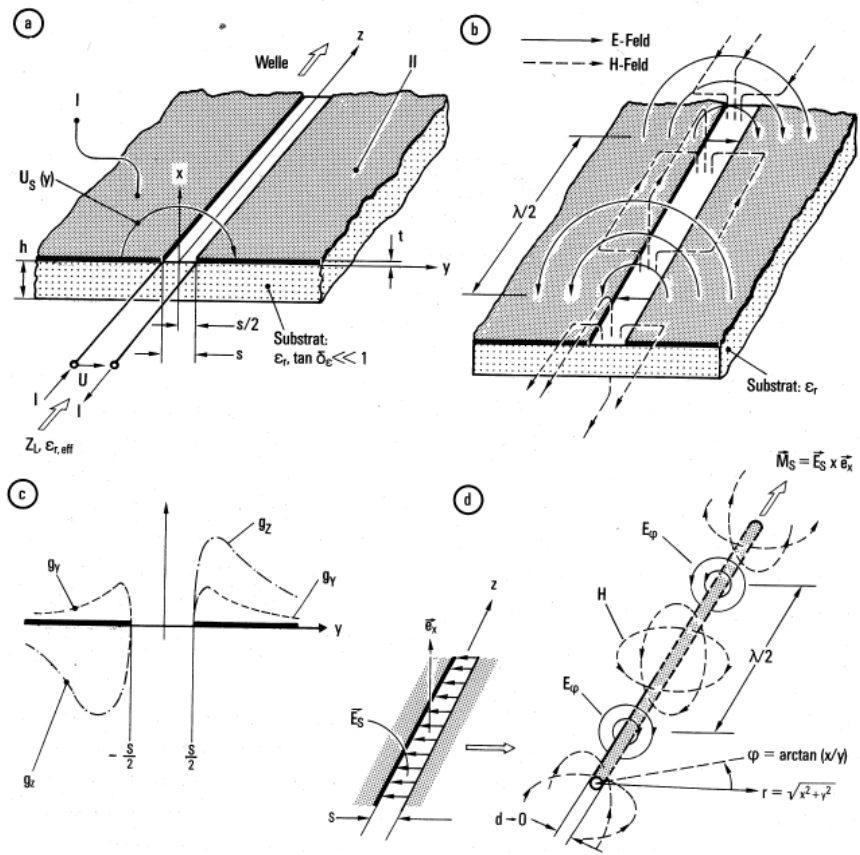

**Fig. 22:** Slotlines a) Mechanical construction, b) Field pattern (TE approximation), c) Longitudinal and transverse current densities, d) Magnetic line current model. Reproduced from Ref. [6] with permission of the author.

Figure 23 shows a broadband (decade bandwidth) pulse inverter. Assuming the upper microstrip to be the input, the signal leaving the circuit on the lower microstrip is inverted since this microstrip ends on the opposite side of the slotline compared to the input. Printed slotlines are also used for broadband pickups in the GHz range, e.g., for stochastic cooling [8].

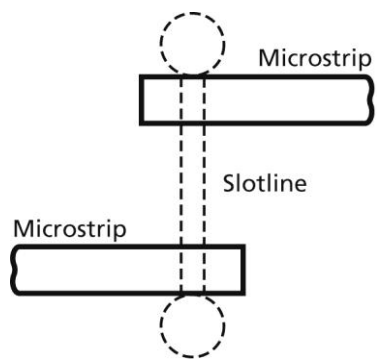

**Fig. 23:** Two microstrip–slotline transitions connected back to back for 180° phase change [7]

## Appendix A: The T-matrix

The T-matrix (transfer matrix), which directly relates the waves on the input and on the output, is defined as [2]

$$\begin{pmatrix} b_1 \\ a_1 \end{pmatrix} = \begin{pmatrix} T_{11} & T_{12} \\ T_{21} & T_{22} \end{pmatrix} \begin{pmatrix} a_2 \\ b_2 \end{pmatrix}. \tag{A1}$$

As the transmission matrix (T-matrix) simply links the in- and outgoing waves in a way different from the S-matrix, one may convert the matrix elements mutually

$$T_{11} = S_{12} - \frac{S_{22}S_{11}}{S_{21}}, \quad T_{12} = \frac{S_{11}}{S_{21}}$$

$$T_{21} = -\frac{S_{22}}{S_{21}}, \quad T_{22} = \frac{1}{S_{21}}. \tag{A2}$$

The T-matrix $\mathbf{T_M}$ of $m$ cascaded 2-ports is given by a matrix multiplication from the 'left' to the 'right' as in Refs. [2, 3]:

$$\mathbf{T_M} = \mathbf{T_1 T_2 \ldots T_m} \tag{A3}$$

There is another definition that takes $a_1$ and $b_1$ as independent variables.

$$\begin{pmatrix} b_2 \\ a_2 \end{pmatrix} = \begin{pmatrix} \tilde{T}_{11} & \tilde{T}_{12} \\ \tilde{T}_{21} & \tilde{T}_{22} \end{pmatrix} \begin{pmatrix} a_1 \\ b_1 \end{pmatrix} \tag{A4}$$

and for this case

$$\tilde{T}_{11} = S_{21} - \frac{S_{22}S_{11}}{S_{12}}, \quad \tilde{T}_{12} = \frac{S_{22}}{S_{12}}$$

$$\tilde{T}_{21} = -\frac{S_{11}}{S_{12}}, \quad \tilde{T}_{22} = \frac{1}{S_{12}}. \tag{A5}$$

Then, for the cascade, we obtain

$$\mathbf{\tilde{T}_M} = \mathbf{\tilde{T}}_m \mathbf{\tilde{T}}_{m-1} \ldots \mathbf{\tilde{T}}_1 \tag{A6}$$

i.e., a matrix multiplication from 'right' to 'left'.

In the following, the definition using Eq. (A1) will be applied. In practice, after having carried out the T-matrix multiplication, one would like to return to S-parameters

$$S_{11} = \frac{T_{12}}{T_{22}}, \quad S_{12} = T_{11} - \frac{T_{12}T_{21}}{T_{22}}$$

$$S_{21} = \frac{1}{T_{22}}, \quad S_{12} = -\frac{T_{21}}{T_{22}}. \tag{A7}$$

For a reciprocal network ($S_{ij} = S_{ji}$) the T-parameters have to meet the condition det $T = 1$

$$T_{11}T_{22} - T_{12}T_{21} = 1. \tag{A8}$$

So far, we have been discussing the properties of the 2-port mainly in terms of incident and reflected waves **a** and **b**. A description in voltages and currents is also useful in many cases. Considering the current $I_1$ and $I_2$ as independent variables, the dependent variables $U_1$ and $U_2$ are written as a Z matrix:

$$U_1 = Z_{11}I_1 + Z_{12}I_2 \\ U_2 = Z_{21}I_1 + Z_{22}I_2 \quad \text{or} \quad (U) = (Z) \cdot (I). \tag{A9}$$

where $Z_{11}$ and $Z_{22}$ are the input and output impedance, respectively. When measuring $Z_{11}$, all the other ports have to be open, in contrast to the S-parameter measurement, where matched loads are required.

In an analogous manner, a *Y*-matrix (admittance matrix) can be defined as

$$I_1 = Y_{11}U_1 + Y_{12}U_2 \\ I_2 = Y_{21}U_1 + Y_{22}U_2 \quad \text{or} \quad (I) = (Y) \cdot (U). \tag{A10}$$

Similarly to the S-matrix, the *Z*- and *Y*-matrices are not easy to apply for cascaded 4-poles (2-ports). Thus, the so-called ABCD matrix (or A matrix) has been introduced as a suitable cascaded network description in terms of voltages and currents (Fig. 1):

$$\begin{pmatrix} U_1 \\ I_1 \end{pmatrix} = \begin{pmatrix} A & B \\ C & D \end{pmatrix} \begin{pmatrix} U_2 \\ -I_2 \end{pmatrix} = \begin{pmatrix} A_{11} & A_{12} \\ A_{21} & A_{22} \end{pmatrix} \begin{pmatrix} U_2 \\ -I_2 \end{pmatrix}. \tag{A11}$$

With the direction of $I_2$ chosen in Fig. 1 a minus sign appears for $I_2$ of a first 4-pole becomes $I_1$ in the next one.

It can be shown that the ABCD-matrix of two or more cascaded 4-poles becomes the matrix product of the individual ABCD-matrices [3]:

$$\begin{pmatrix} A & B \\ C & D \end{pmatrix}_K = \begin{pmatrix} A & B \\ C & D \end{pmatrix}_1 \begin{pmatrix} A & B \\ C & D \end{pmatrix}_2 \cdots \begin{pmatrix} A & B \\ C & D \end{pmatrix}_k. \tag{A12}$$

In practice, the normalized ABCD matrix is usually applied. It has dimensionless elements only and is obtained by dividing *B* by $Z_0$ the reference impedance, and multiplying *C* with $Z_0$. For example, the impedance Z (Fig. 1) with $Z_G = Z_L = Z_0$ would have the normalized ABCD matrix [3, 4]

$$\begin{pmatrix} A & B \\ C & D \end{pmatrix}_N = \begin{pmatrix} 1 & Z/Z_0 \\ 0 & 1 \end{pmatrix}.$$

The elements of the S-matrix are related as

$$S_{11} = \frac{A+B-C-D}{A+B+C+D}, \quad S_{12} = \frac{2 \det A}{A+B+C+D}$$

$$S_{12} = \frac{2}{A+B+C+D}, \quad S_{22} = \frac{-A+B-C+D}{A+B+C+D}, \tag{A13}$$

to the elements normalized of the ABCD matrix. Furthermore, the H matrix (hybrid) and G (inverse hybrid) will be mentioned as they are very useful for certain 2-port interconnections [3].

$$U_1 = H_{11}I_1 + H_{12}U_2 \\ I_2 = H_{21}I_1 + H_{22}U_2 \quad \text{or} \quad \begin{pmatrix} U_1 \\ I_2 \end{pmatrix} = (H) \cdot \begin{pmatrix} I_1 \\ U_2 \end{pmatrix} \tag{A14}$$

and

$$I_1 = G_{11}U_1 + G_{12}I_2$$
$$U_1 = G_{12}U_1 + G_{22}I_2$$

or $\begin{pmatrix} I_1 \\ U_2 \end{pmatrix} = (G) \cdot \begin{bmatrix} U_1 \\ I_2 \end{bmatrix}$ . (A15)

All these different matrix forms may appear rather confusing, but they are applied, in particular, in computer codes for RF and microwave network evaluation. As an example, in Fig. A.1, the four basic possibilities of interconnecting 2-ports (besides the cascade) are shown. In simple cases, one may work with S-matrices directly, eliminating the unknown waves at the connecting points by rearranging the S-parameter equations.

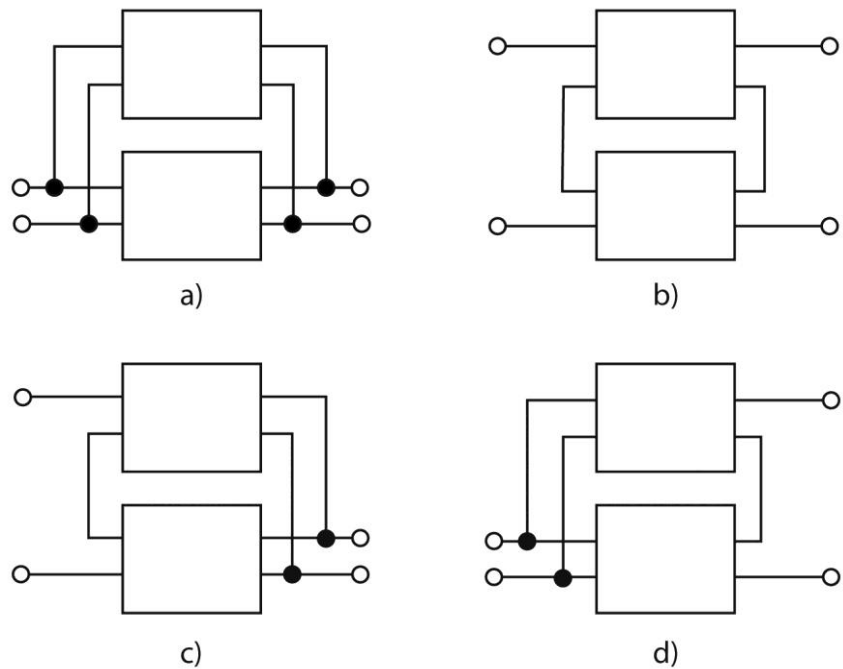

**Fig. A.1:** Basic interconnections of 2-ports [1]: a) Parallel–parallel connection add Y-matrix;  b) Series–series connection add Z-matrix;  c) Series–parallel connection add H matrix;  d) Parallel–series connection add G matrix [3].

Figure A.2 shows ABCD-, S- and T-matrices (reproduced with the permission of the publisher [3]).

| Element | ABCD matrix | S matrix | T matrix |
|---|---|---|---|
| 1. A transmission line section | $\begin{bmatrix} \text{Ch} & Z\,\text{Sh} \\ \dfrac{\text{Sh}}{Z} & \text{Ch} \end{bmatrix}$ | $\dfrac{1}{D_s}\begin{bmatrix} (Z^2 - Z_0^2)\text{Sh} & 2ZZ_0 \\ 2ZZ_0 & (Z^2 - Z_0^2)\text{Sh} \end{bmatrix}$ | $\begin{bmatrix} \text{Ch} - \dfrac{Z^2 + Z_0^2}{2ZZ_0}\text{Sh} & \dfrac{Z^2 - Z_0^2}{2ZZ_0}\text{Sh} \\ -\dfrac{Z^2 - Z_0^2}{2ZZ_0}\text{Sh} & \text{Ch} + \dfrac{Z^2 + Z_0^2}{2ZZ_0}\text{Sh} \end{bmatrix}$ |

where Sh = $\sinh\gamma\ell$,  Ch = $\cosh\gamma\ell$ and $D_s = 2ZZ_0\,\text{Ch} + (Z^2 + Z_0^2)\,\text{Sh}$

| | | | | |
|---|---|---|---|---|
| 2. A series impedance 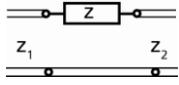 | $\begin{bmatrix} 1 & Z \\ 0 & 1 \end{bmatrix}$ | $\dfrac{1}{D_s}\begin{bmatrix} Z+Z_2-Z_1 & 2\sqrt{Z_1Z_2} \\ 2\sqrt{Z_1Z_2} & Z+Z_1-Z_2 \end{bmatrix}$ | $\dfrac{1}{D_t}\begin{bmatrix} Z_1+Z_2-Z & Z_2-Z_1+Z \\ Z_2-Z_1-Z & Z_1+Z_2-Z \end{bmatrix}$ | |

where $D_s = Z + Z_1 + Z_2$ and $D_t = 2\sqrt{Z_1Z_2}$

| | | | | |
|---|---|---|---|---|
| 3. A shunt admittance 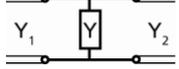 | $\begin{bmatrix} 1 & 0 \\ Y & 1 \end{bmatrix}$ | $\dfrac{1}{D_s}\begin{bmatrix} Y_1-Y_2-Y & \sqrt{2Y_1Y_2} \\ 2\sqrt{Y_1Y_2} & Y_2-Y_1-Y \end{bmatrix}$ | $\dfrac{1}{D_t}\begin{bmatrix} Y_1+Y_2-Y & Y_1-Y_2-Y \\ Y_1-Y_2+Y & Y_1+Y_2+Y \end{bmatrix}$ | |

where $D_s = Y + Y_1 + Y_2$ and $D_t = 2\sqrt{Y_1Y_2}$

| | | | | |
|---|---|---|---|---|
| 4. A shunt-connected open-ended stub 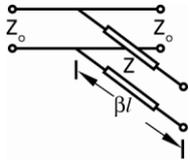 | $\begin{bmatrix} -1 & 0 \\ \dfrac{1}{jZT} & 1 \end{bmatrix}$ | $\dfrac{1}{D_s}\begin{bmatrix} -1 & D_s-1 \\ D_s-1 & -1 \end{bmatrix}$ | $\begin{bmatrix} 1-\dfrac{Z_0}{2Z}T & -j\dfrac{Z_0}{2Z}T \\ j\dfrac{Z_0}{2Z}T & 1+j\dfrac{Z_0}{2Z}T \end{bmatrix}$ | |

where $T = \tan\beta\ell$ and $D_s = 1 + 2jZT/Z_0$

| | | | | |
|---|---|---|---|---|
| 5. A shunt-connected short-circuited stub 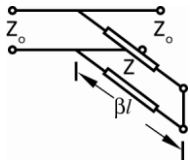 | $\begin{bmatrix} 1 & 0 \\ \dfrac{jT}{Z} & 1 \end{bmatrix}$ | $\dfrac{1}{D_s}\begin{bmatrix} 1 & D_s+1 \\ D_s+1 & 1 \end{bmatrix}$ | $\begin{bmatrix} 1+j\dfrac{Z_0}{2ZT} & j\dfrac{Z_0}{2ZT} \\ -j\dfrac{Z_0}{2ZT} & 1-j\dfrac{Z_0}{2ZT} \end{bmatrix}$ | |

where $T = \tan\beta\ell$ and $D_s = -1 + 2jZ/(Z_0T)$

| | | | | |
|---|---|---|---|---|
| 6. An ideal transformer 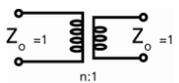 | $\begin{bmatrix} n & 0 \\ 0 & 1/n \end{bmatrix}$ | $\dfrac{1}{n^2+1}\begin{bmatrix} n^2-1 & 2n \\ 2n & 1-n^2 \end{bmatrix}$ | $\dfrac{1}{2n}\begin{bmatrix} n^2+1 & n^2-1 \\ n^2-1 & n^2+1 \end{bmatrix}$ | |
| 7. π-network 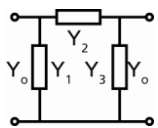 | $\begin{bmatrix} 1+\dfrac{Y_2}{Y_3} & \dfrac{1}{Y_3} \\ \dfrac{D}{Y_3} & 1+\dfrac{Y_1}{Y_3} \end{bmatrix}$ | $\dfrac{1}{D_s}\begin{bmatrix} Y_0^2-PY_0-D & 2Y_0Y_3 \\ 2Y_0Y_3 & Y_0^2+PY_0-D \end{bmatrix}$ | $\dfrac{1}{2Y_0Y_3}\begin{bmatrix} -Y_0^2+QY_0-D & Y_0^2-PY_0-D \\ -Y_0^2-PY_0+D & Y_0^2+QY_0+D \end{bmatrix}$ | |

where $D_s = Y_0^2 + QY_0 + D$, $D = Y_1Y_2 + Y_2Y_3 + Y_3Y_1$, $Q = Y_1 + Y_2 + 2Y_3$ and $P = Y_1 - Y_2$

| 8. T-network 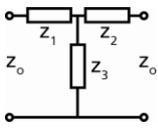 | $\begin{bmatrix} 1+\dfrac{Z_1}{Z_3} & \dfrac{D}{Z_3} \\ \dfrac{1}{Z_3} & 1+\dfrac{Z_2}{Z_3} \end{bmatrix}$ | $\dfrac{1}{D_s}\begin{bmatrix} -Z_0^2 + PZ_0 + D & 2Z_0 Z_3 \\ 2Z_0 Z_3 & -Z_0^2 - PZ_0 + D \end{bmatrix}$ | $\dfrac{1}{2Z_0 Z_3}\begin{bmatrix} -Z_0^2 + QZ_0 - D & -Z_0^2 + PZ_0 + D \\ Z_0^2 + PZ_0 - D & Z_0^2 + QZ_0 + D \end{bmatrix}$ |

where $D_s = Z_0^2 + QZ_0 + D$, $D = Z_1 Z_2 + Z_2 Z_3 + Z_3 Z_1$, $Q = Z_1 + Z_2 + 2Z_3$ and $P = Z_1 - Z_2$

| 9. A transmission line junction 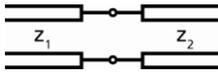 | $\begin{bmatrix} 1 & 0 \\ 0 & 1 \end{bmatrix}$ | $\dfrac{1}{D_s}\begin{bmatrix} Z_2 - Z_1 & 2\sqrt{Z_1 Z_2} \\ 2\sqrt{Z_1 Z_2} & Z_1 - Z_2 \end{bmatrix}$ | $\dfrac{1}{D_s}\begin{bmatrix} Z_1 + Z_2 & Z_2 - Z_1 \\ Z_2 - Z_1 & Z_1 + Z_2 \end{bmatrix}$ |

where $D_s = Z_1 + Z_2$ and $D_t = 2\sqrt{Z_1 Z_2}$

| 10. An α-db attenuator 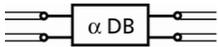 | $\begin{bmatrix} \dfrac{A+B}{2} & Z_0\left(\dfrac{A-B}{2}\right) \\ \dfrac{A-B}{2Z_0} & \dfrac{A+B}{2} \end{bmatrix}$ | $\begin{bmatrix} 0 & B \\ B & 0 \end{bmatrix}$ | $\begin{bmatrix} -A & 0 \\ 0 & A \end{bmatrix}$ |

where $A = 10^{\alpha/20}$ and $B = 1/A$

**Fig. A.2:** (continued) ABCD-, S- and T-matrices for the elements shown

## Appendix B: The signal flow graph (SFG)

The SFG is a graphical representation of a system of linear equations having the general form

$$\mathbf{y} = \mathbf{M}\mathbf{x} + \mathbf{M'}\mathbf{y} \tag{B1}$$

where $\mathbf{M}$ and $\mathbf{M'}$ are square matrices with $n$ rows and columns, $\mathbf{x}$ represents the $n$ independent variables (sources) and $\mathbf{y}$ the $n$ dependent variables. The elements of $\mathbf{M}$ and $\mathbf{M'}$ appear as transmission coefficients of the signal path. When there are no direct signal loops, as is generally the case in practice, the previous equation simplifies to $\mathbf{y} = \mathbf{M}\mathbf{x}$, which is equivalent to the usual S-parameter definition

$$\mathbf{b} = \mathbf{S}\mathbf{a} \ . \tag{B2}$$

The SFG can be drawn as a directed graph. Each wave $a_i$ and $b_i$ is represented by a node, each arrow stands for an S-parameter (Fig. B.1).

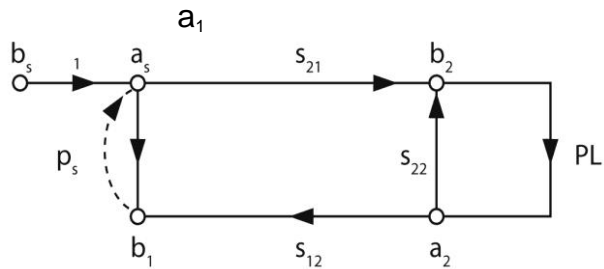

**Fig. B.1:** A 2-port with a non-matched load

For general problems the SFG can be solved by applying Mason's rule. For not too complicated circuits, a more intuitive way is to simplify step-by-step the SFG by applying the following three rules (Fig. B.2):

1. Add the signal of parallel branches
2. Multiply the signals of cascaded branches
3. Resolve loops

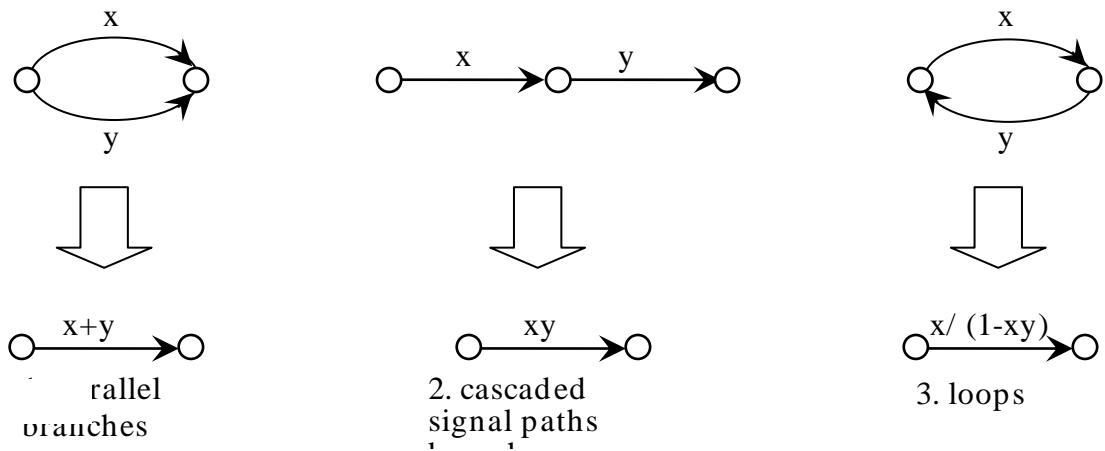

**Fig. B.2:** The three rules for simplifying signal flow charts

Care has to be taken applying the third rule, since loops can be transformed to forward and backward oriented branches. No signal paths should be interrupted when resolving loops.

**Examples**

We are looking for the input reflection coefficient $b_1/a_1$ of a two-port with a non-matched load $\rho_L$ and a matched generator (source $\rho_S = 0$), see Fig. B.1.

The loop at port 2 involving $S_{22}$ and $\rho_L$ can be resolved, given a branch from $b_2$ to $a_2$ with the signal $\Gamma_L *(1-\Gamma_L*S_{22})$. Applying the cascading rule and the parallel branch rule then yields

$$\frac{b_1}{a_1} = S_{11} + S_{21} \frac{\rho_L}{1-S_{22}\rho_L} S_{12} \;. \tag{B3}$$

As a more complicated example one may add a mismatch to the source ($\rho_S$ = dashed line in Fig. B.1) and ask for $b_1/b_S$.

As before, first the loop consisting of $S_{22}$ and $\rho_L$ can be resolved. Then the signal path via $b_2$ and $a_2$ is added to $S_{11}$, yielding a loop with $\rho_S$. Finally one obtains

$$\frac{b_1}{b_S} = \frac{\left(S_{11} + S_{21}S_{12}\rho_L \frac{1}{1-\rho_L S_{22}}\right)}{1-\left(S_{11} + S_{21}S_{12}\rho_L \frac{1}{1-\rho_L S_{22}}\right)\rho_S} \;. \tag{B4}$$

The same results would have been found applying Mason's rule on this problem.

As we have seen in this rather easy configuration, the SFG is a convenient tool for the analysis of *simple* circuits [9, 10]. For more complex networks there is a considerable risk that a signal path may be overlooked and the analysis soon becomes complicated. When applied to S-matrices, the solution may sometimes be read directly from the diagram. The SFG is also a useful way to gain insight into other networks, such as feedback systems. But with the availability of powerful computer codes using the matrix formulations, the need to use the SFG has been reduced.

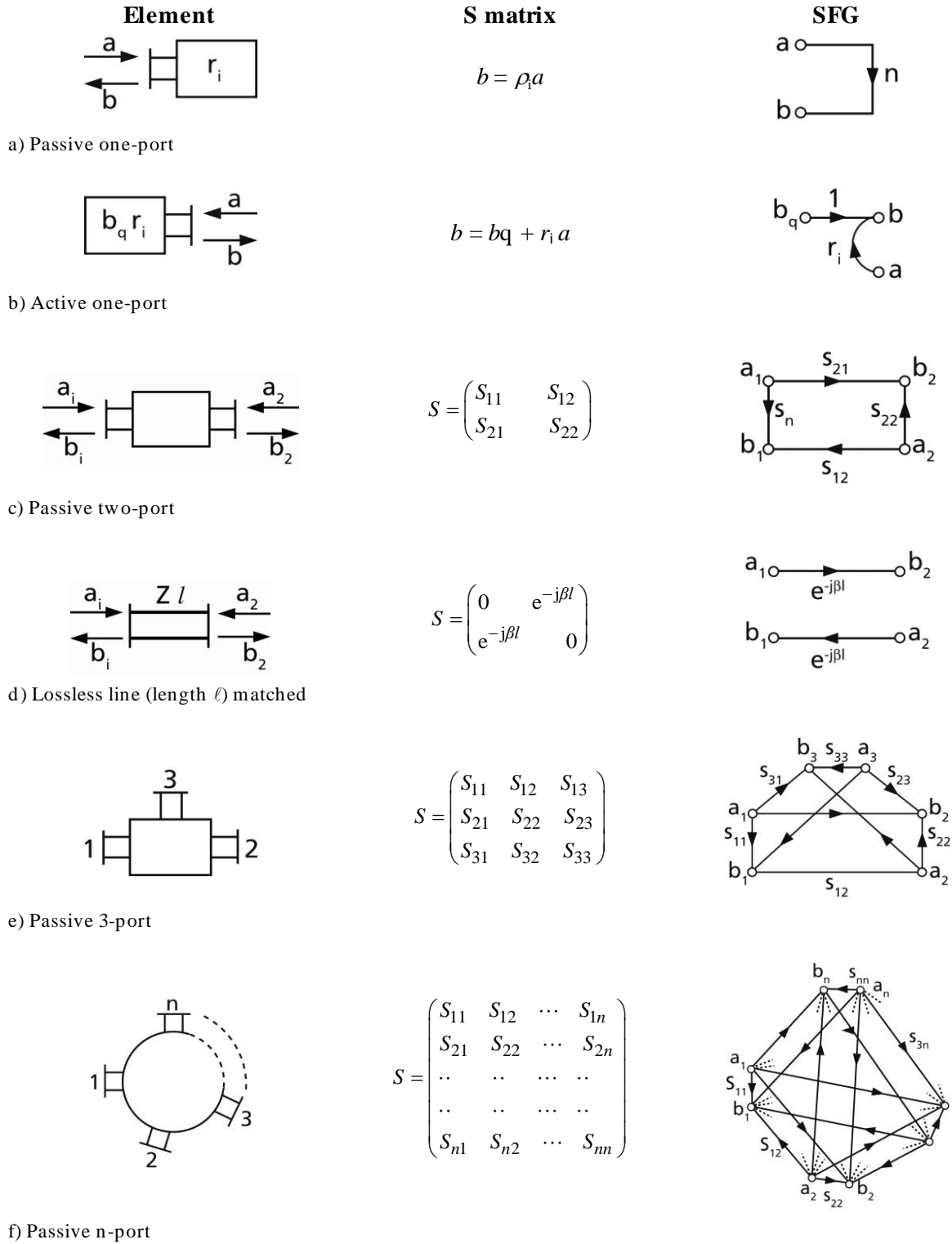

Fig. B.3: SFG and S matrices of different multiports (reproduced from Ref. [10] with the permission of the publisher)

The purpose of the SFG is to visualize physical relations and to provide a solution algorithm of Eq. (B4) by applying a few rather simple rules:

1. The SFG has a number of points (nodes) each representing a single wave $a_i$ or $b_i$.
2. Nodes are connected by branches (arrows), each representing one S-parameter and indicating direction.
3. A node may be the beginning or the end of a branch (arrow).
4. Nodes showing no branches pointing towards them are *source nodes*. All other nodes are *dependent signal nodes*.
5. Each node signal represents the sum of the signals carried by all branches entering it.
6. The transmission coefficients of parallel signal paths are to be added.
7. The transmission coefficients of cascaded signal paths are to be multiplied.
8. An SFG is feedback-loop free if a numbering of all nodes can be found such that every branch points from a node of lower number towards one of higher number.
9. A first-order loop is the product of branch transmissions, starting from a node and going along the arrows back to that node without touching the same node more than once. A second-order loop is the product of two non-touching first-order loops, and an $n^{th}$-order loop is the product of any n non-touching first-order loops.
10. An elementary loop with the transmission coefficient $S$ beginning and ending at a node $N$ may be replaced by a branch $(1-S)^{-1}$ between two nodes $N_1$ and $N_2$, going from $N_1$ to $N_2$. $N_1$ has all signals (branches) previously entering $N$, and $N_2$ is linked to all signals previously leaving from $N$.

In order to determined the ratio $T$ of a dependent to an independent variable the so-called 'non-touching loop rule', also known as Mason's rule, may be applied [11]

$$T = \frac{P_1\left[1-\sum L(1)^{(1)} + \sum L(2)^{(1)} - \cdots\right] + P_2\left[1-\sum L(1)^{(2)} \cdots\right]}{1-\sum L(1)+\sum L(2)-\sum L(3)+\cdots} \quad (B5)$$

where:

- $P_n$ are the different signal paths between the source and the dependent variable.
- $\Sigma L(1)^{(1)}$ represents the sum of all first-order loops not touching path 1, and $\Sigma L(2)^{(1)}$ is the sum of all second-order loop not touching path 1.
- Analogously $\Sigma L(1)^{(2)}$ is the sum of all first-order loops in path 2.
- The expressions $\Sigma L(1)$, $\Sigma L(2)$ etc. in the denominator are the sums of all first-, second-, etc. order loops in the network considered.

**Examples**

We are looking for the input reflection coefficient of a e-port with a non-matched load $\rho_L$ and a matched generator (source) ($\rho_S = 0$) to start with. $\rho_L$, $\rho_S$ are often written as $\Gamma_L$, $\Gamma_S$.

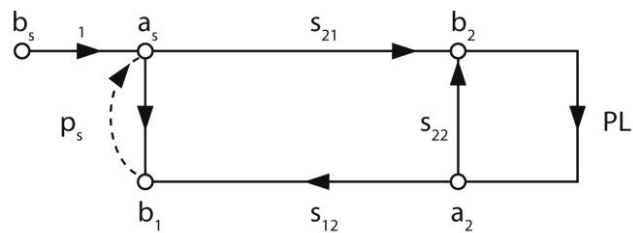

**Fig. B.4:** 2-port with non-matched load

By reading directly from the SFG (Fig. B.4) we obtain

$$\frac{b_1}{a_1} = S_{11} + S_{21}\frac{\rho_L}{1-S_{22}\rho_L}S_{12} \tag{B6}$$

or by formally applying Mason's rule in Eq. (B.5)

$$\frac{b_1}{a_1} = \frac{S_{11}(1-S_{22}\rho_L)+S_{21}\rho_L S_{12}}{1-S_{22}\rho_L}. \tag{B7}$$

As a more complicated example one may add a mismatch to the source ($\rho_S$ = dashed line in Fig. B.4) and ask for $b_1/b_s$

$$\frac{b_1}{b_s} = \frac{S_{11}(1-S_{22}\rho_S)+S_{21}\rho_S S_{12}}{1-(S_{11}\rho_S+S_{22}\rho_L+S_{12}\rho_L S_{21}\rho_S)+S_{11}\rho_{22}S_{22}\rho_L}. \tag{B8}$$